\documentclass[aps,prl,reprint,superscriptaddress,floatfix]{revtex4-1}

\usepackage{slashed}
\usepackage{amsmath}
\usepackage{graphicx,bm}
\usepackage{color}
\usepackage{hyperref}
\usepackage[squaren]{SIunits}

\usepackage[dvipsnames]{xcolor}

\usepackage{verbatim}

\usepackage{grffile} 

\graphicspath{{./images/}}

\let\vec\mathbf

\begin{document}

\title{Towards high photon density for Compton Scattering by spectral chirp} 

\author{M.~A. Valialshchikov}
\email[]{maksim.valialshchikov@skoltech.ru}
\affiliation{Skolkovo Institute of Science and Technology, Moscow 143026, Russia}

\author{D. Seipt}
\affiliation{Helmholtz Institute Jena, Fröbelstieg 3, 07743 Jena, Germany}
\affiliation{GSI Helmholtzzentrum für Schwerionenforschung GmbH, Planckstrasse 1, 64291 Darmstadt, Germany}

\author{V.~Yu. Kharin}
\affiliation{Sechenov First Moscow State Medical University (Sechenov University), Moscow 119435, Russia}

\author{S.~G. Rykovanov}
\email[]{s.rykovanov@skoltech.ru}
\affiliation{Skolkovo Institute of Science and Technology, Moscow 143026, Russia}

\date{\today}


\begin{abstract}
    
Scattering of intense laser pulses on high-energy electron beams allows one to produce a large number of X and $\gamma$ rays. For temporally pulsed lasers the resulting spectra is broadband which severely limits practical applications. One could use linearly chirped laser pulses to compensate that broadening. We show for laser pulses chirped in the spectral domain that there is the optimal chirp parameter at which the spectra has the brightest peak. Additionally, we use catastrophe theory to analytically find this optimal chirp value.
    
\end{abstract}

\maketitle


The scattering of intense laser pulses on high-energy electron beams is a well-established method for generating X and $\gamma-$radiation with application in medicine, ultrafast radiography and nuclear physics \cite{carman1996tunl, bertozzi2008nuclear, albert2010characterization, albert2011design, quiter2011transmission, geddes2015compact}. Recent developments in both compact powerful laser systems and compact laser-plasma based accelerators (LPAs) \cite{geddes2015compact, faure2004laser, blumenfeld2007energy, Nedorezov2004UFN, rykovanov2014quasi, nedorezov2021nuclear} has increased interest in compact Compton photon sources. Small intensities of an incident laser pulse lead to meager photon yields. Increasing laser intensity helps to boost photon yields but also brings nonlinear effects into play, i.e. the spectrum is red-shifted and high harmonics are generated. For temporally pulsed lasers it also leads to a significant spectral ponderomotive broadening \cite{hartemann1996spectral, hartemann2013nonlinear, rykovanov2016controlling, heinzl2010beam, seipt2011nonlinear} which severely limits practical applications of such source. As a result, a lot of research was done to search for methods to compensate or avoid such ponderomotive broadening. For example, it was proposed to use laser pulses with flat-top profiles \cite{hartemann1996spectral} or laser pulse chirping when laser frequency is a nonlinear function of time following the change of laser pulse envelope \cite{rykovanov2016controlling, terzic2014narrow, ghebregziabher2013spectral, seipt2015narrowband}. Recently, it was shown that it is enough to use only linear chirp to significantly reduce ponderomotive broadening \cite{kharin2018higher, seipt2019optimizing}. Moreover, it was proposed to use laser pulses with time-varying polarisation to produce narrowband harmonics in Compton spectra \cite{valialshchikov2021narrow, valialshchikov2021polarisation}. Recently, it was demonstrated that one could use catastrophe theory to analytically study the photon yield enhancement for the case of a laser pulse linearly chirped in the time domain \cite{kharin2018higher}. However, such laser pulses are quite challenging to realize experimentally. Alternative approaches to produce X-rays with low bandwidth include using a travelling-wave setup that allows the overlap of laser and electron beams much longer than Rayleigh length \cite{debus2010traveling, jochmann2013high} or instead of modifying the laser pulse, introduce an additional laser beam co-propagating with the electron bunch \cite{lv2022high}. 

In this paper, we show using catastrophe theory and numerical simulations that using linearly chirped laser pulses in the spectral domain significantly increases the peak photon number compared to the non-chirped nonlinear Compton spectrum. Additionally, we determine analytically the location of optimal chirp parameter at which the photon peak is the highest. Throughout the paper we use natural units $\hbar = c = 1$ while the space-time and energy variables are rescaled by the incident laser frequency $\omega_{0}: x \omega_{0} \rightarrow x,\: \omega/\omega_{0} \rightarrow \omega$. Dimensionless laser pulse amplitude is given by $a_0 = e A/m$, where $e,m$ are the absolute value of electron charge and electron mass respectively. Also, for calculations we use the classical framework since for the parameters of interest the quantum parameter for head-on collision $\chi = 2 a_0 \gamma \omega_{0} / m \ll 1$, where $\gamma$ is the electron gamma factor, and classical description is sufficient.


In the previous work \cite{kharin2018higher}, a laser pulse chirped in the time domain was considered. To move closer to more realistic experimental scenarios
we consider here optimized chirping in the spectral domain.
The laser pulses are modelled as follows,
\begin{equation}
    \tilde A(\omega) = \sqrt{2\pi} a_0 \tau \exp \left (-\frac{\tau^2}{2} (\omega - \omega_0)^2 (1 - i\beta) \right),
\end{equation}
where $\tau$ is the Fourier-limited pulse duration (i.e. with vanishing spectral phase), and second order spectral phase parameter $\beta$ controls the amount of linear chirp.

Such a laser pulse has constant spectrum $|\tilde A(\omega)|^2$ for different chirp parameters. After performing the Fourier transformation one could see that the amplitude and duration of a chirped laser pulse in the time domain depends on parameter $\beta$:
\begin{align}
    &A(\phi) = a_{\text{eff}} \exp\left(-\frac{\phi^2}{2 \tau_{\text{eff}}^2}\right) \exp\left(i(\phi + \frac{\beta \phi^2}{2 \tau_{\text{eff}}^2} + \phi_0)\right),\\ 
    &a_{\text{eff}} = \frac{a_0}{(1 + \beta^2)^{1/4}}, \: \tau_{\text{eff}} = \tau \sqrt{1 + \beta^2},\:
    \omega_L(\phi) = 1 + \frac{\beta \phi}{\tau^2_{\text{eff}}}\nonumber,
\end{align}
where $\phi_0$ is a constant phase shift that depends on $\beta$.

We work in the frame of reference where an electron was initially at rest, $p = (m,0,0,0)$, and results in the laboratory frame are obtained via Lorentz transformation to the frame where electron was initially counter-propagating the laser pulse moving in $+z$ direction. 

The radiation emitted by an electron is given by the scattering integral \cite{jackson1999classical}
\begin{equation}
    \frac{d^2 I}{d\omega d\Omega} = \kappa \frac{\omega^2}{4\pi^2}\left| \int_{-\infty}^{\infty} d\phi \:\: \vec{n} \times [\vec{n} \times \vec{u}] \: e^{i \omega (\phi + z - \vec{n}\cdot \vec{r})} \right|^2 \mathrm{,}
    \label{spec_integral}
\end{equation}
$\kappa = e^2 \omega_0$, $\vec{n}$ is the direction of observation, $\vec{u}, \vec{r}$ are the vector part of electron's 4-velocity and coordinate respectively. We will denote the expression under the modulus by $M: \frac{d^2 I}{d\omega d\Omega} = \kappa \frac{\omega^2}{4\pi^2} |M|^2$. The number of emitted photons are calculated by $\frac{d^2 N_{ph}}{d\omega d\Omega} = \alpha \frac{\omega}{4 \pi^2} |M|^2$, where $\alpha$ is the fine structure constant. 

To transform nonlinear oscillating parts in the exponent into a sum over harmonics, we use Jacobi-Anger expansion. After the transformation for a slowly varying laser pulse envelope one could obtain the following expression
\begin{align}
    M &= \sum_{n=1}^{+\infty} \int_{-\infty}^{\infty}
    \! d\phi \: B_n(\phi; \omega, \beta) \nonumber\\ &\times \exp \left( i\int_{0}^{\phi} \omega + \omega(1 - \cos\theta)\frac{\mathbf{a}^2(\xi)}{4} - n \omega_L(\xi) d\xi \right),
    \label{eq:amplitude}
\end{align}
where $\vec{A}(\phi) = (a_x(\phi) \cos \psi_L(\phi), a_y(\phi) \sin \psi_L(\phi), 0)$, $\psi_L(\phi) = \int_0^{\phi} \omega_L(\xi) d\xi$, $\vec{a}(\phi) = (a_x(\phi), a_y(\phi), 0)$, $\theta$ is the scattering angle and $n$ is the harmonic number.

If the amplitudes $B$ are slowly-varying functions of $\phi$, we could use the stationary phase approximation to estimate the values of integrals (\ref{eq:amplitude}). The stationarity condition determines so-called ray surfaces in the parameter space $(\omega, \theta, \phi)$ for $n$-th harmonic
\begin{align}
    \Phi'(\phi) = \omega + \omega(1 - \cos\theta)\frac{\mathbf{a}^2(\phi)}{4} - n \omega_L(\phi) = 0.
    \label{eq:phi_1}
\end{align}

Ray surfaces from Eq.~(\ref{eq:phi_1}) in the parameter space $(k_x, k_z, \phi) = (\omega \sin \theta, \omega \cos \theta, \phi)$ for a fixed value of $\phi$ determine a set of ellipses in $(k_x, k_z)$ plane for each harmonic \cite{kharin2018higher}. The conditions for fold and cusp singularity (i.e. higher order stationarity) are given by $\Phi''(\phi) = \Phi'''(\phi) = 0$
\begin{align}
    \omega(1 - \cos\theta) \frac{(\mathbf{a}^2)'(\phi)}{4} &= n \omega'_L(\phi) \label{eq:phi_2},\\
    (\mathbf{a}^2)'' &= 0.\label{eq:phi_3}
\end{align}

From Eqs.~(\ref{eq:phi_1})-(\ref{eq:phi_2}) we
obtain expressions for scattered frequency and folds location
\begin{align}
    \omega(\phi, \theta) &= \frac{n \omega_L(\phi)}{1 + (1 - \cos\theta)\mathbf{a}^2(\phi)/4},\label{eq:w}\\
    \cos\theta (\phi) &= 1 - \frac{4\beta}{\omega_L(\phi) [\mathbf{a}^2]'(\phi) \tau^2_{\text{eff}} - \beta \mathbf{a}^2(\phi)}.\label{eq:theta}
\end{align}
Equation (\ref{eq:w}) follows from the stationary phase condition and determines just regular relation between frequency and angle for a chirped pulse. Equation (\ref{eq:theta}) follows from fold condition (\ref{eq:phi_2}) where for frequency $\omega$ we used relation (\ref{eq:w}). For a fixed value of $\phi$ Eqs.~(\ref{eq:w})-(\ref{eq:theta}) give us frequency and angle (if it exists) at which folds are located.

From Eq.~(\ref{eq:phi_3}) we obtain the point at which the folds coincide ($\phi_c = -\tau_{\text{eff}}/\sqrt{2}$) and that determines the cusp singularity.

From now on, we consider circularly-polarized laser pulses $a_x = a_y$ and the main Compton line $n=1$. We would like to find the optimal $\beta$ parameter for fixed $a_0, \: \tau$ for which the spectra would have the brightest maximum. Eq.~(\ref{eq:theta}) shows that for narrowband and collimated emission the cusp angle should lie close to the axis. Imposing an additional constraint $\theta_c = \pi$ on Eq.~(\ref{eq:theta}), we obtain an equation for $\beta_c$
\begin{align}
    \beta_c \left( 1 + \frac{2 a_0^2}{\sqrt{1 + \beta_c^2}} e^{-1/2} \right) = \sqrt{2} \tau a_0^2 e^{-1/2}
\end{align}
where the exponential multiplier is due to the Gaussian temporal envelope and which could be solved either numerically or perturbatively for effectively large $a_0^2\tau$: $\beta_c\simeq \sqrt{2e^{-1}}a_0^2\tau -2\sqrt{2}e^{-1} \frac{a_0^4\tau}{\sqrt{1+2e^{-1}a_0^4\tau^2}} $.
This equation gives us the chirp value $\beta_c$ at which the cusp would lie on-axis. The corresponding frequency is given by $\omega_c = 1 - \frac{\sqrt{2} \beta_c}{\tau \sqrt{1 + \beta_c^2}}$.

One could notice that emission is \emph{not} brightest exactly at the cups. The values of the integral Eq.~(\ref{eq:amplitude}) near the cusp are determined by the Pearcey diffraction pattern. Thus, we do not have to demand that the cusp is on axis but rather that the maximum of the Pearcey pattern is located on axis.
Taylor-expanding the integral under the exponential in the amplitude $M$ around the cusp point $\phi_c$ we obtain the Pearcey integral $Pe$ (see Supplemental Material)
\begin{align}
    \frac{d^2 N_{ph}}{d\omega d\Omega}_{|\theta=\pi} \approx \alpha \frac{\omega}{8\pi^2} \frac{a_0^2 e^{-1/4}}{\sqrt{1 + \beta^2}} \left( \frac{6\sqrt{2} \tau_{\text{eff}}^3}{\omega a_{\text{eff}}^2} \right)^{1/2} |Pe(x,y)|^2,
    \label{eq:Pearcey_max}
\end{align}
where $x,\: y$ are functions of $\omega,\: \beta$. Knowing the values $x_*,\: y_* \approx -2.16, 0$ at which $|Pe(x,y)|$ achieves its maximum, we obtain two equations $x(\omega, \beta) = x_*,\: y(\omega, \beta)=y_*$ which gives us the optimal $(\omega_*, \beta_*)$ pair found by analyzing the maximum of the Pearcey integral.

It turns out that even with Pearcey maximum the optimal chirp parameter is not found because the prefactor $B$ also depends on $\omega, \beta$ that could not be neglected in our case due to their variation in the region of interest (see Equation \ref{eq:Pearcey_max}): $\frac{d^2 N_{ph}}{d\omega d\Omega} \sim \sqrt{\omega (1 + \beta^2)} |Pe(x,y)|^2$. To improve our theoretical prediction of optimal chirp, we Taylor-expanded this expression around the Pearcey maximum $(\omega_*, \beta_*)$ up to the 2nd order assuming that the actual optimal pair lies close to the Pearcey pair (see Supplemental Material) and obtain corrected optimal values $(\omega_T, \beta_T)$. In the text we would call this procedure ``Taylor prefactor correction''.

\begin{figure}[b]
    \centering
    \includegraphics[width=\linewidth]{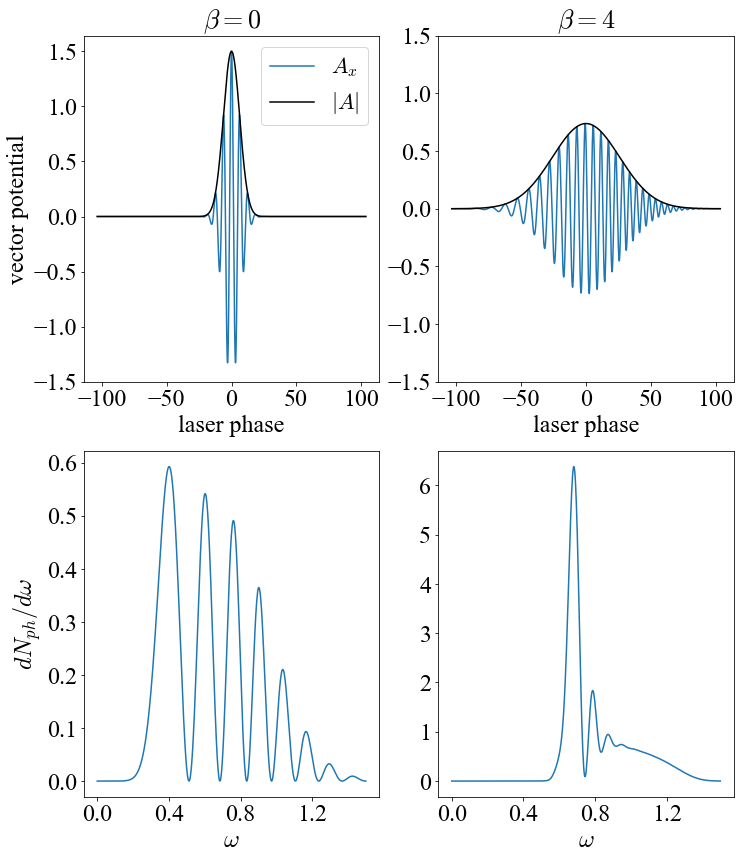}
    \caption{(Top) Vector potential of a linearly chirped laser pulse ($a_0 = 1.5, \tau=2\pi$) for different chirp parameters $\beta = 0, 4$. (Bottom) Corresponding backscattered spectra.}
    \label{fig:A_spectra}
\end{figure}

Figure \ref{fig:A_spectra} illustrates how varying the linear chirp parameter $\beta$ affects the incident laser pulse and subsequently the emission spectra. When the chirp is increased two effects come into play. On the one hand, stronger chirping produces less intense and longer laser pulses for which the backscattered spectra is less broad and therefore more bright because the incident laser spectrum is constant. On the other hand, increasing the chirp moves the cusp point closer to axis which, up to some point, also increases the peak spectrum value. For instance, we could see that changing $\beta$ from 0 to 4 in the laser pulse with $a_0=1.5,\: \tau=2\pi$ leads to a narrower spectra with smaller number of subpeaks and a weak pedestal is created. For large $\beta$ the cusp point vanishes and increasing $\beta$ even further leads to larger spectral pedestal along with smaller spectral peak until it reaches $\beta \gg 1$ limit. 

\begin{figure}
    \centering
    \includegraphics[width=0.9\linewidth]{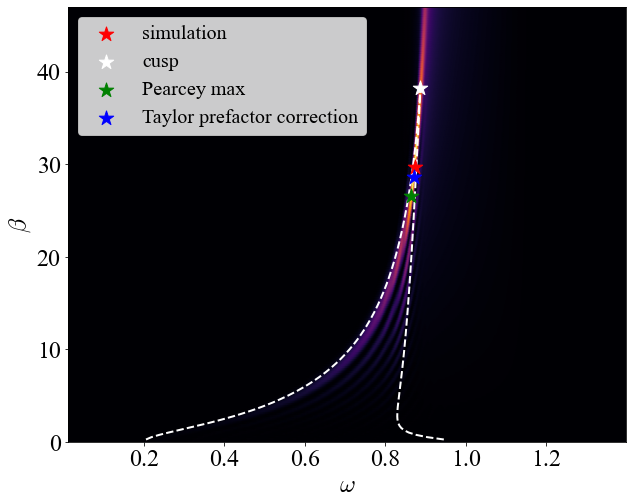}
    \caption{Backscattered spectra for different $\beta$ parameters and $a_0=2, \tau=4\pi$. Dashed white line shows folds terminating at the cusp (white star). Red star shows the maximal peak found from simulations, green and blue stars show the location of the maximal peak obtained from different analytic methods.}
    \label{fig:folds}
\end{figure}

Let us illustrate how folds and cusp look like in our problem setting. Figure \ref{fig:folds} shows backscattered spectra from a laser pulse ($a_0=2, \tau=4\pi$) for different $\beta$ parameters. Due to the dependence of laser amplitude and duration on chirp, the folds profile significantly differs from the original article \cite{kharin2018higher}. For larger laser intensities and duration the maximal peak moves to higher frequencies and chirps and enters the region where the emission frequency is almost constant. That is why even Pearcey approximation (when we do not take into account the prefactor that depends on $\omega, \beta$) gives satisfactory results.

\begin{figure}
    \centering
    \includegraphics[width=0.9\linewidth]{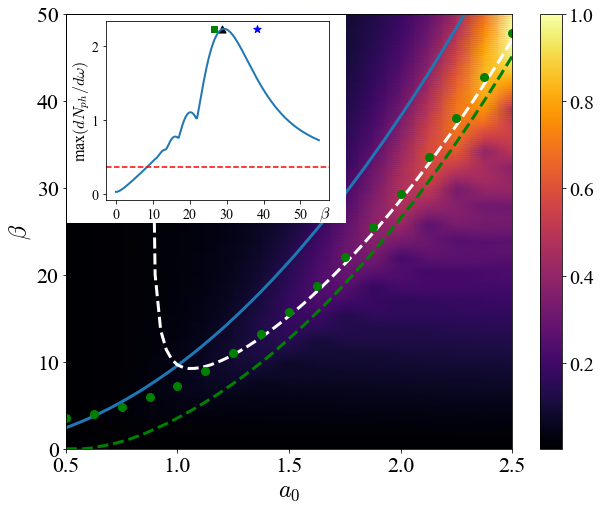}

    \caption{Normalized peak of the differential number of emitted photons (colormap) as a function of laser amplitude $a_0$ and chirp parameter $\beta$ for $\tau=4\pi$. Green dots show the optimal $\beta$ at which the peak has the highest value for a fixed $a_0$, solid blue line shows $\beta$ as a function of $a_0$ obtained by solving the system of cusp-on-axis equations, dashed green line shows optimal $\beta(a_0)$ obtained by Pearcey approximation of the scattering integral, while the white dashed line shows $\beta(a_0)$ obtained by Taylor prefactor correction procedure to Pearcey approximation. One could see that from $a_0 > 1$ there is a very good agreement between Taylor prefactor correction and actual numerical optima. (Inset) Slice of the colormap at $a_0=2$, black triangle stands for the prediction of optimal $\beta$ from Taylor procedure (blue and green symbols stand for cusp-on-axis and Pearcey prediction respectively), red dashed line shows the limit $\beta \gg 1$ for given $a_0, \tau$.}
    \label{fig:2d_beta_a0}
\end{figure}

To obtain numerical data we fixed laser pulse amplitude and duration ($a_0, \tau$) and calculated the results on a linear grid over $\beta$ parameters. Numerical optimal $\beta$ value corresponds to the maximal photon peak observed from the simulation results on the grid. Figure \ref{fig:2d_beta_a0} shows the colormap for the normalized peak of the differential number of emitted photons as a function of $(a_0, \beta)$ for $\tau = 4\pi$. We could see that for a fixed $a_0$ there is the optimal $\beta$ value (which is greater for larger $a_0, \tau$) at which the photon peak is maximal. Our goal is to estimate analytically this optimal value. One could see that solving the system of cusp-on-axis equations overshoots the target value and the agreement is generally not very good. With Pearcey approximation the analytical prediction is much better for larger $a_0$ but still not good enough. Finally, Taylor prefactor correction procedure gives a very satisfactory analytical prediction. The same could be seen on the inset slice. Similar results were found for different laser pulse duration (for $\tau = 2\pi, \: 6\pi$ see Supplemental Material) although for larger duration the analytical prediction starts to work from less intense values of $a_0$.

\begin{figure}
    \centering
    \includegraphics[width=0.9\linewidth]{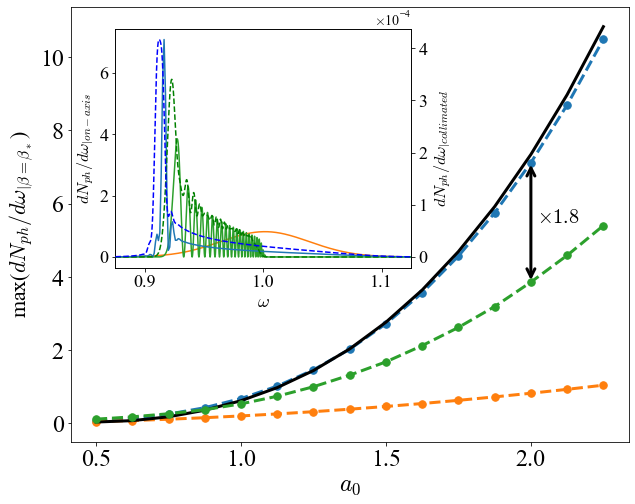}
    
    \caption{Peak of the differential number of emitted photons estimated at optimal chirp value $\beta_*$ as a function of $a_0$ for $\tau = 6\pi$. Blue line shows the peak values estimated from numerical simulations, black solid line shows the analytical prediction from Pearcey approximation, green line shows the peak of the unchirped spectra for $a_{\text{eff}}, \tau_{\text{eff}}$ estimated at optimal chirp value $\beta_*$ and orange line shows linear Compton limit for $\beta \gg 1$. For example, for $a_0 = 2$ there is almost double enhancement in the peak value compared to non-chirped case and over 8 time enhancement compared to the limit case $\beta \gg 1$. (Inset) Solid lines show on-axis spectra (left y-axis): blue shows optimally chirped case, green shows unchirped nonlinear Compton spectra for $a_{\text{eff}}, \tau_{\text{eff}}$ estimated at optimal chirp value $\beta_*$, orange shows the linear Compton limit case for $a_0, \tau$. Blue and green dashed lines (right y-axis) show the optimally chirped spectra and nonlinear Compton spectra collimated over $\theta_{c}=0.1/\gamma$.}
    \label{fig:Nmax_optimal_linear}
\end{figure}

To demonstrate the benefit of using chirped laser pulses we compared the peak photon spectrum values with the un-chirped case. Figure \ref{fig:Nmax_optimal_linear} illustrates that using optimally chirped laser pulses increases the peak of the differential number of emitted photons compared to un-chirped nonlinear Compton spectrum and limit case for $\beta \gg 1$. For larger $a_0$ the enhancement is more noticeable. On the inset figure we could see that the optimally chirped spectra has a narrow and bright single peak on top of a weaker pedestal while unchirped nonlinear Compton spectrum presents typical broadband interference substructure. After collimation over a small angle $\theta_c = 0.1/\gamma$ the spectrum remains narrow but for larger collimaiton angles it would be significantly broader. It would be interesting to have a look at real values corresponding to a found optimally chirped laser pulse. For example, for $a_0 = 2, \: \tau = 6\pi$ the optimal chirp $\beta_{opt} \approx 48$. For a laser with $\omega_0 = 1.55$ eV the duration would be $\tilde \tau = 8$ fs (bandwidth $\sim 1/\tau \approx 0.05$) and group delay dispersion (GDD) $\beta^{(2)} = \beta_{opt} \tilde \tau^2 \approx 3072$ $\text{fs}^2$. 

In this article we considered the case of a laser pulse chirped in the spectral domain. In the temporal domain the amplitude and duration of such laser pulse depend on chirp parameter. We showed that there is an optimal chirp at which the backscattered spectrum has the brightest photon peak which results from the interplay of two effects. First, when we increase chirp the laser energy remains constant while the backscattered spectrum is shrunk towards $\omega \sim 1$ which increases the maximal peak. Second, larger chip leads to the cusp moving closer to axis which makes the spectra brighter as well until the cusp goes too far and the spectral peak begins to fall down. Also, we discussed the analytic approach (and some improvements to it) to predict the optimal chirp value and showed that for relatively large laser amplitudes and laser duration the given approach works really well and one does not need to perform costly linear scans to find the values of interest. Also, chirping the laser pulse gives a significant improvement over an unchirped case in terms of spectrum narrowness and brightness. Finally, these results could help to move closer towards the experimental realization.

M.V. would like to thank P. Petriakova for inspiration and great discussions. The work was supported by the Russian Science Foundation (Grant No. 22-22-01031) \cite{grant}. The code to reproduce the results from the article could be found in \cite{github}. The results presented in this paper are based on work performed before Feb 24th 2022.


%

\end{document}


\maketitle

\tableofcontents

\section{Fourier transform}
Laser impulse chirped in spectral domain
\begin{equation}
    \tilde A(\omega) = \sqrt{2\pi} a_0 \tau \exp \left (-\frac{\tau^2}{2} (\omega - \omega_0)^2 (1 - i\beta) \right),
\end{equation}
Performing Fourier transform and considering $\omega_0 = 1$, we obtain the expression in time domain
\begin{align}
    &A(\phi) = a_{\text{eff}} \exp\left(-\frac{\phi^2}{2 \tau_{\text{eff}}^2}\right) \exp\left(i(\phi + \frac{\beta \phi^2}{2 \tau_{\text{eff}}^2} + \phi_0)\right),\\ 
    &a_{\text{eff}} = \frac{a_0}{(1 + \beta^2)^{1/4}}, \: \tau_{\text{eff}} = \tau \sqrt{1 + \beta^2},\:
    \omega_L(\phi) = 1 + \frac{\beta \phi}{\tau^2_{\text{eff}}}\nonumber,\\
    &\phi_0 = \arctan \frac{\beta}{1 + \sqrt{1 + \beta^2}} \nonumber,\\
    &A_x(\phi) = a_{\text{eff}} \exp\left(-\frac{\phi^2}{2 \tau_{\text{eff}}^2}\right) \cos\left(\phi + \frac{\beta \phi^2}{2 \tau_{\text{eff}}^2} + \phi_0\right),\nonumber\\
    &A_y(\phi) = a_{\text{eff}} \exp\left(-\frac{\phi^2}{2 \tau_{\text{eff}}^2}\right) \sin\left(\phi + \frac{\beta \phi^2}{2 \tau_{\text{eff}}^2} + \phi_0\right).\nonumber
\end{align}

\newpage

\section{Pearcey approximation}
In the vicinity of the cusp the backscattered spectrum is given by
\begin{align}
    \frac{d^2 N_{ph}}{d\omega d\Omega}_{|\theta=\pi} \approx \alpha \frac{\omega}{8\pi^2} \frac{a_0^2 e^{-1/4}}{\sqrt{1 + \beta^2}} \left( \frac{6\sqrt{2} \tau_{\text{eff}}^3}{\omega a_{\text{eff}}^2} \right)^{1/2} |Pe(x,y)|^2,
    \label{eq:Pearcey_max}
\end{align}
where $Pe(x,y) = \int_{-\infty}^{\infty}\exp(i(t^4 + x t^2 + y t))dt$, $(x,y)$ are functions of $(\omega, \beta)$. It could be shown by Taylor expansion of the integral under the exponential in the amplitude $M$ around the cusp position $\phi_c$ and transforming the integral to the standard Pearcey form.

The Pearcey arguments $(x,y)$ have the following form (here we kept the radiation angle $\theta$ but for calculations one should assume $\theta=\pi$ so the Pearcey maximum would lie on-axis)
\begin{align*}
    &\tau_{\text{eff}} = \tau \sqrt{1 + \beta^2},\:
    \tilde a^2 = 2 a_{\text{eff}}^2 e^{-1/2} = \frac{2 a_0^2}{\sqrt{1 + \beta^2}} e^{-1/2},\\
    &x = - \left( \frac{3 \sqrt{2} \tau_{\text{eff}}}{\omega(1 - \cos\theta) \tilde a^2} \right)^{1/2} \left[ \frac{1}{2\sqrt{2}} \omega (1 - \cos\theta) \tilde a^2 - \frac{\beta}{\tau_{\text{eff}}} \right],\\
    &y = - \left( \frac{12 \sqrt{2} \tau_{\text{eff}}^3}{\omega (1 - \cos\theta) \tilde a^2} \right)^{1/4} \left[ \omega (1 + \frac{1}{4}(1 - \cos\theta) \tilde a^2) -1 + \frac{\beta}{\sqrt{2}\tau_{\text{eff}}} \right].
\end{align*}

It is known that the Pearcey diffraction pattern $|Pe(x,y)|$ achieves its maximum at $(x_*, y_*) \approx (-2.16, 0) = (-C_\mathrm{max}, 0)$. Assuming $x(\omega, \beta) = x_*,\: y(\omega, \beta) = y_*$ gives us two equations from which we could find the optimal pair $(\omega_*, \beta_*)$ obtained by analyzing Pearcey maximum. The aforementioned equations are given by
\begin{align*}
    \omega_* &= \frac{1 - \frac{\beta_*}{\sqrt{2} \tau_{\text{eff,*}}}}{1 + \frac{1}{4}(1 - \cos\theta) \tilde a^2_*},\\
    \frac{1}{2\sqrt{2}}(1 - \cos\theta) \tilde a^2_* - \frac{\beta_*}{\tau_{\text{eff,*}}} \left( 1 + \frac{1}{2}(1 - \cos\theta) \tilde a^2_* \right) &= C_{\text{max}} \sqrt{\frac{\omega_* (1 - \cos\theta) \tilde a^2_*}{3\sqrt{2}\tau_{\text{eff,*}}}} \left( 1 + \frac{1}{4}(1 - \cos\theta) \tilde a^2_* \right),
\end{align*}
where $a^2_*, \tau_{\text{eff,*}}$ means that the effective amplitude and duration correspond to chirp value $\beta_*$.

Assuming $\theta = \pi$ gives us
\begin{align}
    \omega_* &= \frac{1 - \frac{\beta_*}{\sqrt{2} \tau_{\text{eff,*}}}}{1 + \frac{1}{2} \tilde a^2_*},\\
    \frac{1}{\sqrt{2}} \tilde a^2_* - \frac{\beta_*}{\tau_{\text{eff,*}}} \left( 1 + \tilde a^2_* \right) &= C_{\text{max}} \sqrt{\frac{2 \omega_* \tilde a^2_*}{3\sqrt{2}\tau_{\text{eff,*}}}} \left( 1 + \frac{1}{2} \tilde a^2_* \right) \label{eq_sup:beta_Pe}.
\end{align}
which needs to be solved for $\beta_*$ and $\omega_*$.

Equation (\ref{eq_sup:beta_Pe}) with explicit dependence on $\beta_*$ is given by
\begin{align*}
    \frac{\sqrt{2} a_0^2 e^{-1/2}}{\sqrt{1 + \beta^2_*}} - \frac{\beta_*}{\tau \sqrt{1 + \beta^2_*}} \left(1 + \frac{2 a_0^2 e^{-1/2}}{\sqrt{1 + \beta^2_*}}\right) = C_{\text{max}} \left[ \frac{2 \sqrt{2} a_0^2 e^{-1/2}}{3 \tau (1 + \beta^2_*)} \left( 1 - \frac{\beta_*}{\sqrt{2} \tau \sqrt{1 + \beta^2_*}} \right) \left( 1 + \frac{a_0^2 e^{-1/2}}{\sqrt{1 + \beta^2_*}} \right) \right]^{1/2}
\end{align*}

\section{Taylor prefactor correction procedure}
From Eq. (\ref{eq:Pearcey_max}) it could be seen that the prefactor of $|Pe(x,y)|^2$ depends on $\omega$ and $\beta$ and their variation could not be neglected in certain parameter regions if one wants to find the optimal chirp value. Omitting the multipliers that do not depend on $\omega, \beta$ one obtains
\begin{align*}
    \frac{d^2 N_{ph}}{d\omega d\Omega}_{|\theta=\pi} \sim \sqrt{\omega (1 + \beta^2)} |Pe(x,y)|^2 = g(\omega, \beta) |Pe(x,y)|^2 = f(\omega, \beta).
\end{align*}

To find the spectra maximum we want $\partial f/\partial \omega = \partial f / \partial \beta = 0$. To find the solution we Taylor expand $f(\omega, \beta)$ around the Pearcey maximum $(\omega_*, \beta_*)$ up to the 2nd order
\begin{align*}
    f(\omega, \beta) \approx f_{|*} + \frac{\partial f}{\partial \omega}_{|*} (\omega - \omega_*) + \frac{\partial f}{\partial \beta}_{|*} (\beta - \beta_*) + \frac{1}{2} \frac{\partial^2 f}{\partial \omega^2}_{|*} (\omega - \omega_*)^2 \\+ \frac{\partial^2 f}{\partial \omega \partial \beta}_{|*} (\omega - \omega_*) (\beta - \beta_*) + \frac{1}{2} \frac{\partial^2 f}{\partial \beta^2}_{|*} (\beta - \beta_*)^2 + ...\:,
\end{align*}
where $\cdot_{|*}$ means that the derivative is evaluated at the point $(\omega_*, \beta_*)$.

Demanding the first derivatives of $f$, Taylor-expanded up to the second order in $(\omega, \beta)$, to be equal to zero ($\partial f/\partial \omega = \partial f / \partial \beta = 0$) gives us the system of two equations from which we could find the corrected value of optimal pair $(\omega_T, \beta_T)$
\begin{align}
    \frac{\partial f}{\partial \omega}_{|*} + \frac{\partial^2 f}{\partial \omega^2}_{|*} (\omega_T - \omega_*) + \frac{\partial^2 f}{\partial \omega \partial \beta}_{|*} (\beta_T - \beta_*) = 0,\\
    \frac{\partial f}{\partial \beta}_{|*} + \frac{\partial^2 f}{\partial \beta^2}_{|*} (\beta_T - \beta_*) + \frac{\partial^2 f}{\partial \omega \partial \beta}_{|*} (\omega_T - \omega_*) = 0.
\end{align}

It is important to note that Taylor expanding helped to define an explicit linear system of equations where the derivatives of $f$ are just given numbers and not functions of $(\omega, \beta)$. Eliminating one variable we obtain the following system
\begin{align}
    \omega_T - \omega_* = - \frac{\frac{\partial f}{\partial \omega}_{|*} + \frac{\partial^2 f}{\partial \omega \partial \beta}_{|*} (\beta_T - \beta_*)}{\frac{\partial^2 f}{\partial \omega^2}_{|*}},\\
    \frac{\partial f}{\partial \beta}_{|*} + \frac{\partial^2 f}{\partial \beta^2}_{|*} (\beta_T - \beta_*) - \frac{\partial^2 f / \partial \omega \partial \beta_{|*}}{\partial^2 f/ \partial \omega^2_{|*}} \left( \frac{\partial f}{\partial \omega}_{|*} + \frac{\partial^2 f}{\partial \omega \partial \beta}_{|*} (\beta_T - \beta_*) \right) = 0.
    \label{eq:Taylor_beta}
\end{align}

To solve Eq.(\ref{eq:Taylor_beta}) one needs to explicitly calculate all derivatives at the Pearcey maximum point. The calculation is quite long, here we mention only some important aspects. Firstly, there are derivatives $\partial |Pe(x,y)|/\partial \omega_{|*}$, $\partial |Pe(x,y)|/\partial \beta_{|*}$ which are equal to zero because they are calculated at the maximum point. Secondly, in our calculation we encounter the second derivatives $\partial^2 |Pe(x,y)|/\partial \omega^2_{|*}$, $\partial^2 |Pe(x,y)|/\partial \beta^2_{|*}$, $\partial^2 |Pe(x,y)|/\partial \omega \partial \beta_{|*}$ which were calculated numerically using finite differences. Other explicit expressions for the derivatives calculation are present in the code (function \textit{calculate$\_$Taylor$\_$correction$\_$to$\_$Pearcey$\_$max}) \cite{github}.

\newpage

\section{Additional figures}
\begin{figure}[h!]
    \centering
    \includegraphics[width=0.8\textwidth]{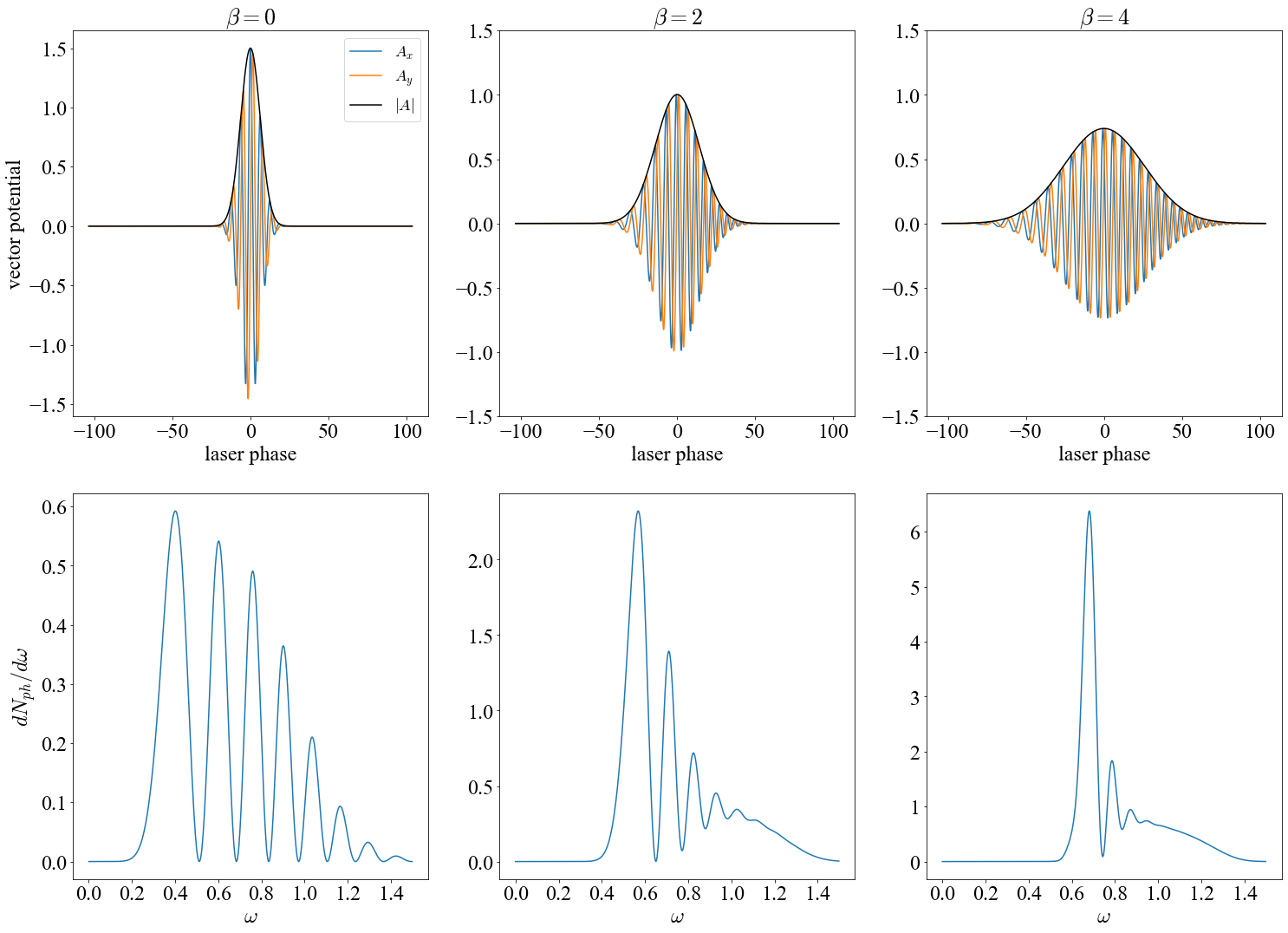}
    \caption{(Top) Vector potential of a linearly chirped laser pulse in spectral domain ($a_0 = 1.5, \tau=2\pi$) for different chirp parameters $\beta = 0, 2, 4$. (Bottom) Corresponding backscattered spectra.}
    \label{fig:A_spectra}
\end{figure}

Figure \ref{fig:A_spectra} illustrates how intensity and duration of a laser pulse chirped in spectral domain dynamically changes when increasing chirp parameter $\beta$. As a result, backscattered spectra also changes: for $\beta=0$ a regular nonlinear Compton spectra with bandlike structure is observed while for larger chirps the spectrum becomes less broad and a pedestal appears.

\begin{figure}
    \centering
    \begin{subfigure}{0.32\textwidth}
        \centering
        \includegraphics[width=\textwidth]{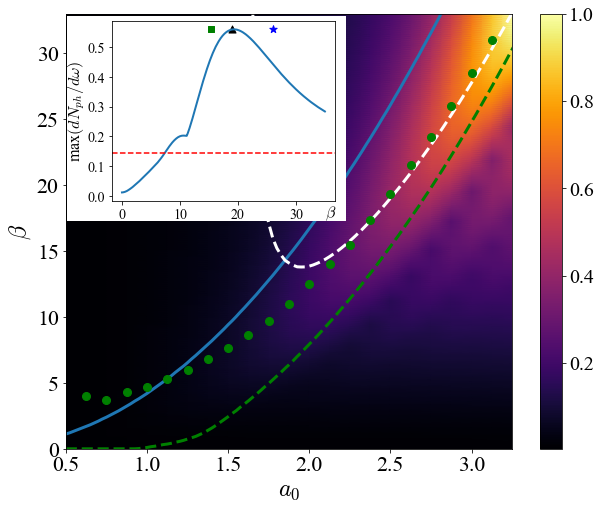}
    \end{subfigure}
    \hfill
    \begin{subfigure}{0.32\textwidth}
        \centering
        \includegraphics[width=\textwidth]{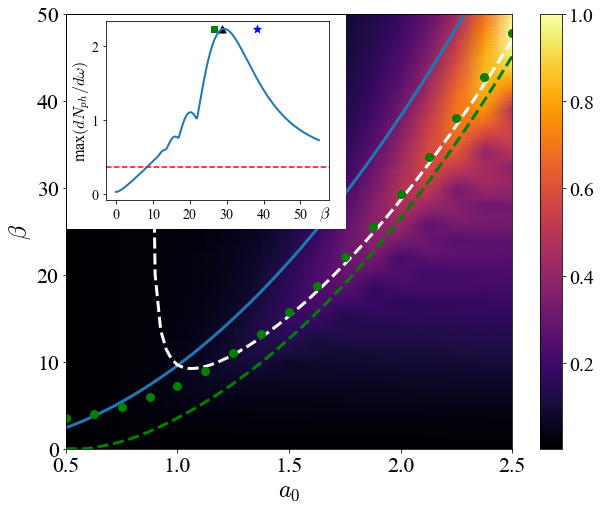}
    \end{subfigure}
    \hfill
    \begin{subfigure}{0.32\textwidth}
        \centering
        \includegraphics[width=\textwidth]{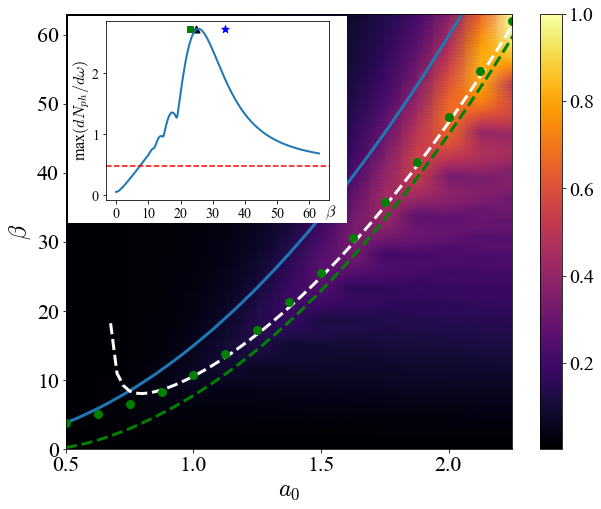}
    \end{subfigure}
    \caption{Normalized peak of the differential number of emitted photons (colormap) as a function of laser amplitude $a_0$ and chirp parameter $\beta$ for (left) $\tau=2\pi$, (middle) $\tau=4\pi$, (right) $\tau=6\pi$. Green dots show the optimal $\beta$ at which the peak has the highest value for a fixed $a_0$, solid blue line shows $\beta$ as a function of $a_0$ obtained by solving the system of cusp-on-axis equations, dashed green line shows optimal $\beta(a_0)$ obtained by Pearcey approximation of scattering integral, white dashed line shows $\beta(a_0)$ obtained by Taylor correction procedure to Pearcey approximation. One could see that from $a_0 > 1$ there is a very good agreement between Taylor correction and actual numerical optima. (Inset) Slice of the colormap at (left) $a_0=2.5$, (middle) $a_0=2$, (right) $a_0=1.5$ black triangle stands for the prediction of optimal $\beta$ from Taylor procedure (blue and green symbols stand for cusp-on-axis and Pearcey prediction respectively), red dashed line shows the limit $\beta \gg 1$ for given $a_0, \tau$.}
    \label{fig:2d_colormap}
\end{figure}

On Figure \ref{fig:2d_colormap} one could see colormaps of normalized peak of the differential number of emitted photons as a function of laser amplitude $a_0$ and chirp $\beta$ for different laser pulse duration: $\tau = 2\pi, 4\pi, 6\pi$. One could see that analytic approximations work ``better'' for larger $\tau$ because in our approximation we considered the prefactors (temporal envelope) to vary slowly. For larger $a_0$ the optimal $\beta$ is larger meaning that the effective pulse duration becomes larger as well that is why for larger $\tau$ the approximation starts to work for smaller $a_0$.

\begin{figure}
    \centering
    \begin{subfigure}{0.32\textwidth}
        \centering
        \includegraphics[width=\textwidth]{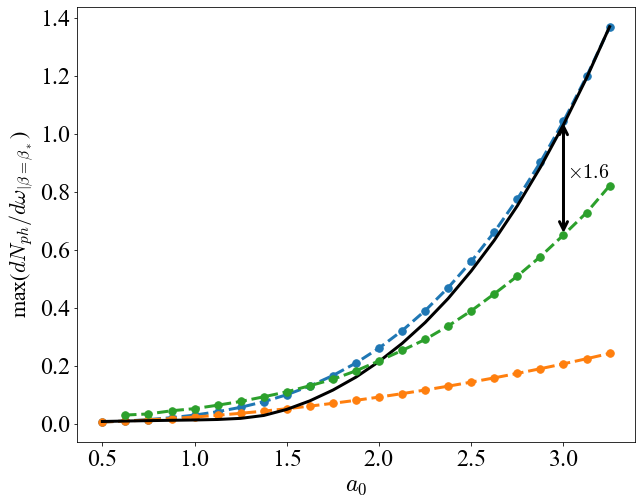}
    \end{subfigure}
    \hfill
    \begin{subfigure}{0.32\textwidth}
        \centering
        \includegraphics[width=\textwidth]{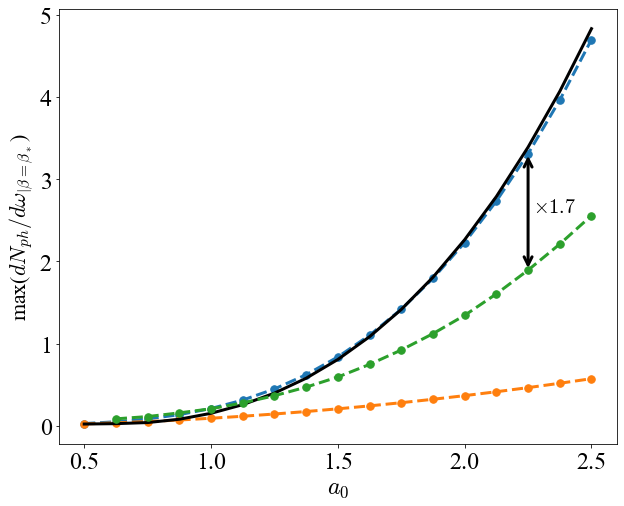}
    \end{subfigure}
    \hfill
    \begin{subfigure}{0.32\textwidth}
        \centering
        \includegraphics[width=\textwidth]{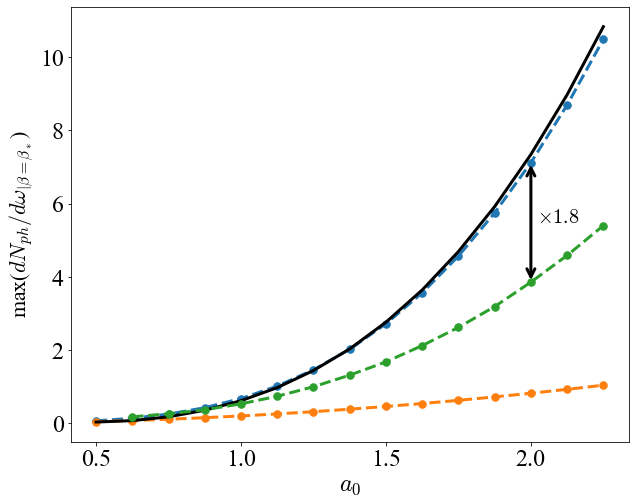}
    \end{subfigure}
    \caption{Peak of the differential number of emitted photons estimated at optimal chirp value $\beta_*$ as a function of $a_0$ for (left) $\tau=2\pi$, (middle) $\tau=4\pi$, (right) $\tau=6\pi$. Blue line shows the peak values estimated from numerical simulations, black solid line shows the analytical prediction from Pearcey approximation, green line shows the peak of the unchirped spectra for $a_{\text{eff}}, \tau_{\text{eff}}$ estimated at optimal chirp value $\beta_*$ and orange line shows linear Compton limit for $\beta \gg 1$.}
    \label{fig:Nmax_optimal}
\end{figure}

Figure \ref{fig:Nmax_optimal} shows the comparison between peak values of differential number of emitted photons for optimally chirped laser pulse, nonlinear Compton spectra for corresponding $a_{\text{eff}}, \tau_{\text{eff}}$ and the linear Compton limit for $\beta \gg 1$. From all cases it could be seen that using optimally chirped laser pulses gives an enhancement.

\begin{figure}
    \centering
    \begin{subfigure}{0.32\textwidth}
        \centering
        \includegraphics[width=\textwidth]{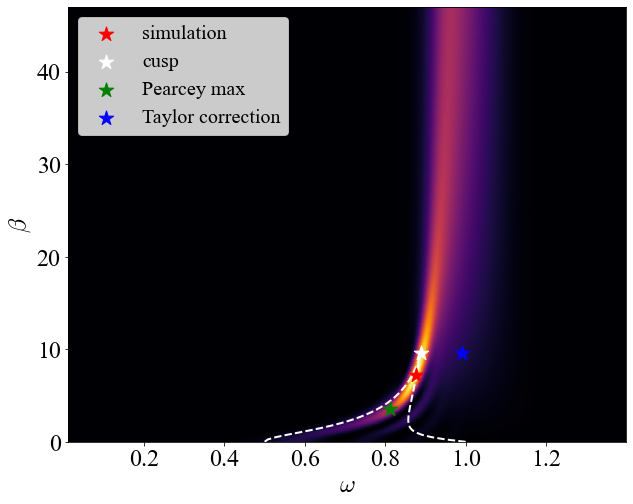}
    \end{subfigure}
    \hfill
    \begin{subfigure}{0.32\textwidth}
        \centering
        \includegraphics[width=\textwidth]{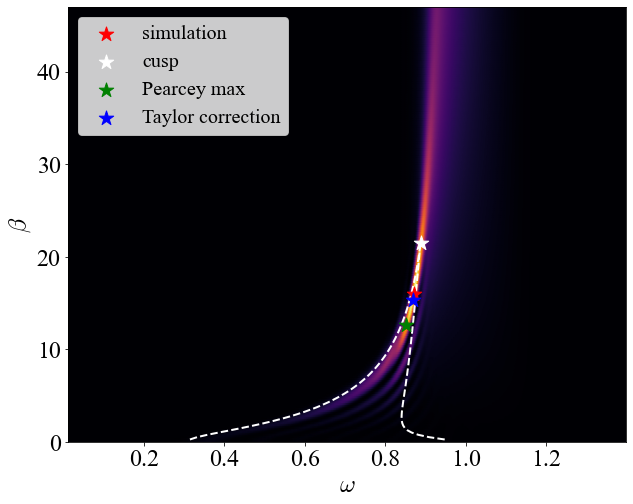}
    \end{subfigure}
    \hfill
    \begin{subfigure}{0.32\textwidth}
        \centering
        \includegraphics[width=\textwidth]{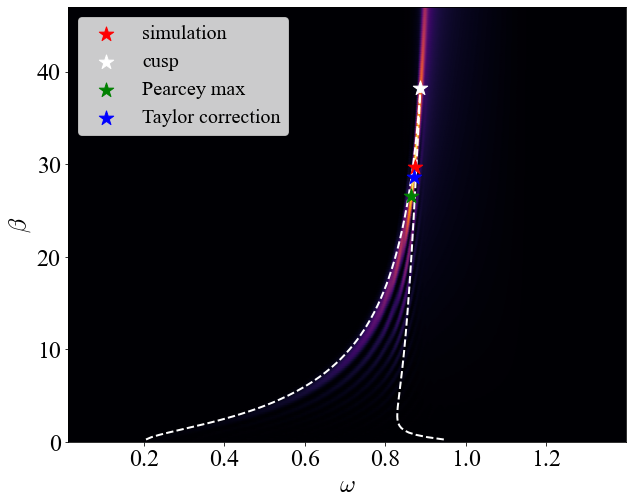}
    \end{subfigure}
    \caption{Backscattered spectra for different $\beta$ parameters and $\tau=4\pi$, (left) $a_0=1$, (middle) $a_0=1.5$, (right) $a_0=2$. Dashed white line shows folds terminating at the cusp (white star). Red star shows the maximal peak found from simulations, green and blue stars show the location of the maximal peak obtained from different analytic methods.}
    \label{fig:folds_cusp}
\end{figure}

Finally, Figure \ref{fig:folds_cusp} demonstrates the backscattered spectra for different chirps $\beta$ for $\tau = 4\pi$ and different intensities $a_0 = 1, 1.5, 2$ with folds terminating at the cusp.

\newpage

\section{Numerical optimization}
In the main text we used linear grid scans to determine the optimal chirp $\beta$ at which the spectrum has the highest peak. In other words, we defined a linear grid over $\beta$ parameters and performed the simulation for each grid point. If we need just to find the optimal parameter without obtaining a nice graph over $\beta$ grid it is much more efficient to use smarter optimization techniques, e.g. Bayesian optimization. For instance, a very straightforward and efficient tool for such task would be Python package \textit{optuna} \cite{optuna}. An example of usage of such efficient optimization applied to our problem setting is shown on Figure given on Github page \cite{github}. It is important to note that \textit{optuna} performed much less simulations to find the optimal value and it is not restricted to grid points which may yield more precise results.

\begin{figure}[h!]
    \centering
    \includegraphics[width=0.6\textwidth]{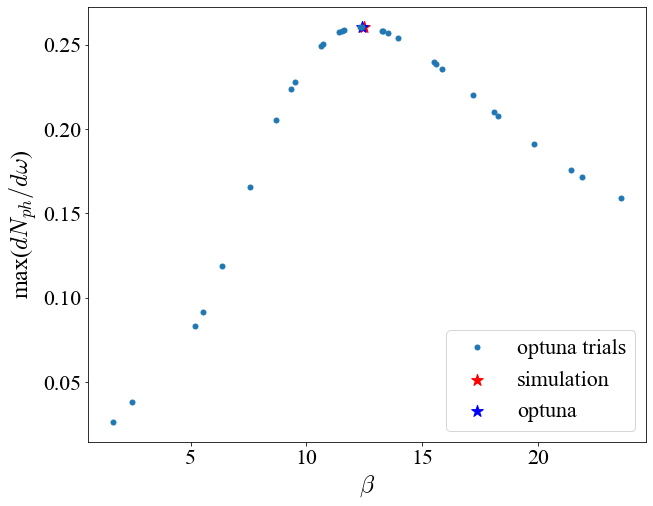}
    \caption{Numerical optimization (\textit{optuna} package) to find optimal chirp $\beta$ for a laser pulse with $a_0 = 2, \: \tau = 2\pi$. \textit{Optuna} found the optimal value after 30 trials (different $\beta$ values) while we performed simulations on a $\beta$ grid with 211 points.}
    \label{fig:optuna}
\end{figure}

\printbibliography